\documentclass[12pt]{article}
\begin{document}
%\selectlanguage{english}
%%%%%%%%%%%%%%%%%%%%%%%%%%%%%%%%%%%%%%%%%%%%%%%%%%%%%%%%%%%%%%%%%%%%%%%%%%
\def\textindent#1{\indent\llap{#1\enspace}\ignorespaces}
\def\cap#1{\textindent{\bf Fig.~#1: }}
\def\simgt{\mathrel{\lower .3ex \rlap{$\sim$}\raise .5ex \hbox{$>$}}}
\def\simlt{\mathrel{\lower .3ex \rlap{$\sim$}\raise .5ex \hbox{$<$}}}
\def\avg#1{{\left\langle#1\right\rangle}}         % average <x>
\def\abs#1{{\left|#1\right|}}                     % absolute value
\def\tdn2n{{\left\langle\left(\delta N\right)^2\right\rangle / \left\langle N\right\rangle}}
\def\na{n_{\mathrm{A}}}
\def\nb{n_{\mathrm{B}}}
\def\xa{x_{\mathrm{A}}}
\def\xb{x_{\mathrm{B}}}
\def\ea{\varepsilon_{\mathrm{A}}}
\def\eb{\varepsilon_{\mathrm{B}}}
\def\ei{\varepsilon_i}
\def\eo{\varepsilon_o}
\def\kb{k_{\mathrm{B}}}
\def\kt{k_{\mathrm{B}}T}
\def\c2x2{c$(2\times 2)$}

\renewcommand{\baselinestretch}{1.3}\small\normalsize
\title{\bf Thermal percolation for interacting monomers adsorbed on square lattices}
\vspace{1mm}
\renewcommand{\baselinestretch}{1.0}\small\normalsize
\author{M.C. Gim\'enez, F. Nieto and A.J. Ramirez-Pastor\thanks{
\underbar{\bf Corresponding author}: A.J. Ramirez-Pastor, 
Departamento de F\'{\i}sica, Universidad Nacional de San Luis, CONICET,
Chacabuco 917, 5700, San Luis, Argentina.
{\bf Email:} antorami@unsl.edu.ar } \\[5mm]
Departamento de F\'{\i}sica, Universidad Nacional de San Luis, \\
 CONICET, Chacabuco 917, 5700 San Luis, Argentina \\
 cecigime@unsl.edu.ar; fnieto@unsl.edu.ar; \\antorami@unsl.edu.ar}
\maketitle

\begin{abstract}
In this paper the percolation of monomers on a square lattice is studied as the
particles interact with either repulsive or attractive energies. By means of a
finite-size scaling analysis, the critical exponents and the scaling collapsing
of the fraction of percolating lattice are found. A phase diagram separating a
percolating from a non-percolating region is determined. The main features of
the phase diagram are discussed in terms of simple considerations related to
the interactions present in the problem. The influence of the phase transitions
occurring in the system is reflected by the phase diagram. In addition, a
scaling treatment maintaining constant the surface coverage and varying the
temperature of the system is performed. In all the considered cases, the
universality class of the model is found to be the same as for the random
percolation model.

\end{abstract}

\noindent Pacs: 64.60.Ak; 68.35.Rh; 68.35.Fx

\newpage

\section{Introduction}

Percolation theory has attracted a great deal of interest in the last few decades and the
activity in the field is still growing \cite{Ham,Stauffer,cardy1,Addler,Sahimi,Zallen,Kirk,
Essam,Hovi,Ziff1,Coniglio,Vale1,Vale2}. 
This is mainly because some aspects of the
percolation process such as the geometrical phase transitions occurring in the system have
gained a particular impetus due to the introduction of techniques such as Monte Carlo (MC)
simulations and series expansions \cite{Essam,Binder}. However, the problem is far from being exhausted.

In fact, most of the studies are devoted to the percolation of molecules that are irreversibly
deposited. In part this is due to the fact that the deposition (or irreversible adsorption)
of particles on solid surfaces is a subject of considerable practical importance. In many
experiments on adhesion of colloidal particles and proteins on solid substrates, the relaxation
time scales are much longer than the times of the formation of the deposit. In such processes,
the temperature of the system does not play any relevant role and it is not considered.
However, in numerous systems of both theoretical and practical importance, where the adsorbed
particles are in thermodynamic equilibrium, the spatial distribution of the adsorbate might
be characterized by using the percolation model \cite{Gao1,Gao2}. 
In these cases, the temperature governs the phase in the system and can be an important 
controlling factor in the percolation process. 
In the simplest case, where repulsive monomers are adsorbed on a square lattice, the
system exhibits a continuous phase transition from a disordered state at high $T$ to a 
doubly-degenerated $c(2 \times 2)$ ordered state at low $T$, with a critical temperature 
$T_c$ which satisfies
$K_c = 1.7637$ \cite{Onsager}, with $K = w/k_BT$, $k_B$ the Boltzmann constant and $w$ 
the interaction energy between nearest-neighbour adatoms. 
On the other hand, the same system with attractively
interacting particles goes through a first-order phase transition. 
It is clear that a percolation
study on the spatial configuration of the adlayer should recognize peculiarities related to the
phase transitions occurring in the adsorbate. Such a study implies the determination of the
critical parameters as a function of the concentration (surface coverage) and temperature. To
the best of our knowledge this study has not been done and it is the main purpose of this paper.

The paper is organized as follows. In section \ref{BD} the model of adsorption of interacting
monomers on a two-dimensional square lattice is presented. The analysis of results obtained
by using finite-size scaling theory is given in section \ref{fss}. In section \ref{PD}, the phase diagram is
discussed along with the basis of a thermal finite-size scaling study. Finally, conclusions are
drawn in section \ref{C}.

\section{Basic definitions}\label{BD}

Let us consider that the substrate is represented by a two-dimensional square lattice of
$M = L \times L$ equivalent adsorption sites, with periodic boundary conditions. 
In order to
describe the system of $N$ monomers adsorbed on $M$ sites at a given temperature $T$, let us
introduce the occupation variable $c_i$ which can take the following values:

\begin{equation}
c_i=\cases{1,& if site i is occupied\cr
           0,& if site i is vacant.}
\end{equation}

Particles can be adsorbed on the substrate with the restriction of at most one adsorbed
particle per site and we consider a nearest-neighbour ($NN$) interaction energy $w$ between
them. 
Under these considerations, the Hamiltonian of the system is given by

\begin{equation}
H = w \sum_{(i,j)'} c_i c_j + \sum_{i}^M \varepsilon_i c_i \label{h}
\end{equation}

\noindent where $(i,j)'$ represents pairs of $NN$ sites and $\varepsilon_i$ is the 
adsorption energy of the sites on the surface. 
In addition, we have taken $\varepsilon_i=\varepsilon=0$ without loss of generality.

For fixed values of surface coverage, $\theta = N/M$, and temperature $T$, the thermodynamic
equilibrium is reached in the canonical ensemble by using a standard Kawasaki algorithm
\cite{kawasaki}. The procedure is as follows. 
An initial arbitrary configuration of $N$ adsorbed monomers
with the desired surface coverage is generated. 
Two sites are randomly selected, and their positions are established. 
Then, an attempt is made to interchange their occupancy state with
probability given by the Metropolis rule \cite{Metropolis}:

\begin{equation}
P = min \left\{1,exp\left( -\Delta H / k_B T\right) \right\}
\end{equation}

\noindent where $\Delta H=H_f -H_i$ is the difference between the Hamiltonian evaluated at the final state
and the one computed at the initial state. A Monte Carlo step (MCS) is achieved when $M$
pair of sites have been tested to change its occupancy state. The time (expressed in units of
MCS) required for equilibration depends on lattice size, temperature and coverage. Typically,
$m = 10^5$ MCSs suffices for smaller lattices containing up to $32 \times 32$ sites in the whole
range of both temperature and coverage. Then, a set of $m = 2 \times 10^3$ samples in thermal
equilibrium is generated by taking configurations separated from each other by $r$ MCSs in
order to avoid possible correlations between the states. In the smallest lattices considered
in this paper, $r = 10^3$ prevents such undesired correlations. The accuracy of the procedure
and the correctness of the algorithm were tested by obtaining the behaviour of different
quantities (for example, the adsorption isotherms (surface coverage, $\theta$, versus the normalized
chemical potential $\mu / k_B T$), the specific heat, the order parameter, etc) and comparing
with the corresponding ones derived from the real space renormalization group approach
(RSRG) \cite{NTU01}.

The central idea of the percolation theory is based on finding the minimum concentration
$\theta$ for which at least a cluster (a group of occupied sites in such a way that each site has at least
one occupied nearest-neighbour site) extends from one side to the opposite one of the system.
This particular value of the concentration rate is named {\it critical concentration} or 
{\it percolation threshold} and determines a phase transition in the system. In this paper, 
the percolation process will be studied under two different perspectives. Namely, samples will 
be prepared for fixed temperature (coverage) and variable coverage (temperature). We call this 
feature {\it coverage percolation} ({\it thermal percolation}).

In the random percolation model, a single site is occupied with probability $p$. The samples
are then generated through irreversible adsorption. In our problem, the occupancy state of
every site strongly depends on $K$. In both cases, for a precise value of concentration, the
percolation threshold of sites, at least one spanning cluster connects the borders of the system
(indeed, there exists a finite probability of finding $n$ ($>1$) spanning 
clusters \cite{aizen,cardy2,shchur1,shchur2}). 
Then, a second-order phase transition appears at such coverage which is characterized by well-defined
critical exponents. It should be emphasized that for $K = 0$ the system becomes uncorrelated
and formally we have the random percolation model.

A study of the finite-size effects allows us to make a reliable extrapolation to the
thermodynamic limit ($L \to \infty$). Details of this study will be given below.

\section{Finite-size scaling}\label{fss}

It is well known that it is a quite difficult matter to analytically determine the value of the
percolation threshold for a given lattice \cite{Stauffer,Sahimi,Zallen,Essam,Hovi}. 
For some special types of lattices,
geometrical considerations enable us to derive their percolation thresholds exactly. Thus,
exact thresholds for the random percolation problem are known for (a) square, triangular
and honeycomb lattices and (b) triangular and Kagome lattice concerning the bond and site
problem, respectively. In both cases, analytical results are obtained when a monomeric
species is considered. For different conditions, i.e. for systems which do not present such a
topological advantage, percolation thresholds have to be estimated numerically by means of
computer simulations.

As the scaling theory predicts \cite{Binder}, the larger the system size to study, the more accurate
the values of the threshold obtained therefrom. Thus, the finite-size scaling theory gives us
the basis to achieve the percolation threshold and the critical exponents of a system with a
reasonable accuracy. 
For this purpose, the probability $R=R^X _L(\theta)$ that a lattice composed of
$L \times L$ elements (sites or bonds) percolates at concentration $\theta$ can 
be defined \cite{Stauffer}. Here, as in \cite{J1,J2}, 
the following definitions can be given according to the meaning of $X$: 
(a) $R^{R (D)} _L(\theta)$ is the probability of finding a rightward (downward) percolating cluster; 
(b) $R^{I} _L(\theta)$ is the probability that we find a cluster which percolates both 
in a rightward {\bf and} in a downward direction; 
(c) $R^{U} _L(\theta)$ is the probability of finding either a rightward {\bf or} a downward percolating
cluster and 
(d) $R^{A} _L(\theta) \equiv \frac{1}{2} \left[R^{R} _L(\theta)+R^{D} _L(\theta) \right] \equiv \frac{1}{2} \left[R^{I} _L(\theta)+R^{U} _L(\theta) \right] $.

The first step for determining the percolation threshold consists in evaluating the effective
threshold $\theta_c(L)$ (the concentration at which the slope of $R^{X} _L(\theta)$ is the largest) 
for a lattice of finite-size $L$. In the MC simulations, $R^{X} _L(\theta)$ is determined for 
each discrete value of $\theta$ according to the considered finite regular lattice \cite{Vale2}. 
Once the temperature is fixed, the next
procedure is followed: (a) the construction of $m$ samples for a given coverage (according
to the scheme presented in section \ref{BD}) and (b) the cluster analysis by using the Hoshen and
Kopelman algorithm \cite{HKA}. In the last step, the number of clusters for each sample, $n_s$, of
size $s$ (a cluster of size $s$ is composed by $s$ connected elements) is determined in order to
verify whether a percolating island exists. This spanning cluster could be determined by
using the criteria $R$, $D$, $I$ or $U$. $m$ runs of two such steps are carried out for obtaining the
number $m^X$ of them for which a percolating cluster of the desired criterion $X$ is found.
Then, $R^X _L(\theta)=m^X/n$ is defined and the procedure is repeated for different values both of $\theta$ and
lattice sizes, $L$ ($L = 32, 48, 64, 80, 96$ and $128$).

In figure $1$, the probabilities $R^{I} _L(\theta)$ (squares), $R^{U} _L(\theta)$ (circles) 
and $R^{A} _L(\theta)$ (triangles) are presented. 
Three different values of $K$ are shown. From a first inspection of the figure (and
from data not shown here for the sake of clarity) it is observed that (a) curves cross each
other in a unique universal point, $R^{X^*}$ , which depends on the criterion $X$ used; (b) those points
do not modify their height for the different $K$’s. This finding indicates, as is expected, that
the universality class of the phase transition involved in the problem is conserved no matter
what the value of $K$ is; (c) those points are located at very well-defined values in the $\theta$-axes,
determining the critical percolation threshold for each $K$; (d) the standard percolation problem
is recovered for $K = 0$ giving a critical coverage $\theta_c=0.5927$ and (e) $\theta_c$ increases (decreases)
for positive (negative) value of $K$. A detailed explanation of this point will be given below.

The second step is the extrapolation of $\theta_c ^X(L)$ towards the limit $L \to \infty$ 
by using the scaling hypothesis. Thus, the correlation length, $\xi$ , can be expressed as

\begin{equation}
\xi \propto \vert \theta -\theta_c \vert ^{-\nu}
\label{cld}
\end{equation}

\noindent where the critical exponent $\nu$ is analytically shown to be equal to $\nu=4/3$ 
in the case of random
percolation \cite{Stauffer,Addler,Sahimi,Zallen}. As $\theta=\theta_c ^X(L)$
the correlation length reaches the linear dimension $L$ of the lattice. 
Thus, we have

\begin{equation}
\theta_c ^X(L)=\theta_c(\infty)+ A^X L^{-1/\nu}
\label{extrapolation}
\end{equation}

\noindent where $A^X$ is a non-universal constant. 
Figure $2$ shows the extrapolation towards the thermodynamic limit of $\theta_c ^X(L)$ according to equation 
(\ref{extrapolation}) for different values of $K$ as indicated.
This figure lends support to the assertion given by equation (\ref{extrapolation}): 
(a) all the curves (different criteria) are well correlated by a linear function, 
(b) they have a quite similar value for the ordinate in the $L\to \infty$ 
and (c) the fitting determines a different value of the constant
$A$ depending on the type of criterion used. 
It is also important to note that $\theta_c ^A(L)$ gives an
almost perfect horizontal line which is a great advantage of the method because it does not
require precise values of critical exponents in the process of estimating percolation thresholds.
The maximum of the differences between $\vert \theta_c ^I(\infty) - \theta_c ^A(\infty) \vert$ and  
$\vert \theta_c ^U(\infty) - \theta_c ^A(\infty) \vert$ give the error bar for each determination 
of $\theta_c$.

The scaling law hypothesis also predicts the collapsing of the curves $R^X _L (\theta)$ 
when they are plotted as a function of a reduced variable $u= \left( \theta - \theta_c \right) L^{1/\nu}$:

\begin{equation}
R^X=\overline{R^X}\left( \left( \theta - \theta_c \right) L^{1/\nu} \right),
\label{functionR}
\end{equation}

\noindent $\overline{R^X}(u)$ being with the scaling function. 
Thus, $\overline{R^X}$ is a universal function with respect to the variable $u$. 
In figure $3 (a)$, as an illustration, we plot $R^X_L$ as a function of $u$ for $K = 0$. 
This gives an additional indication for the numerical value of the critical exponent $\nu$. 
As is clearly seen from this analysis, the problem belongs to the same universality class 
of random percolation regardless of the value of $K$ considered.

The same procedure described above can be realized for different values of $K$ as is shown
in figure $3 (b)$ just for $K = 0$, $1$ and $2.5$. Thus, for a given value of $K$, all the curves used
in the experiment (for different values of $L$) collapse into a universal curve according to the
theoretical prediction. However, $\overline{R^X}$ is not only a function of $\theta$ and $L$ 
but also of $K$. As can be
seen, the collapsing function is different for each value of $K$ considered. This fact determines
that the scaling function $\overline{R^X}$ is not a universal function with respect to the variable $K$ 
(each value of $K$ is represented by using a different type of line as indicated).

In order to determine the dependence of $\overline{R^X}$ with $K$, the main features of the collapsing
data have to be considered in the range of $K$ between $-2$ and $3$. As can be seen, the curves
become steeper upon increasing the value of $K$. In fact, the derivative of the universal function
$\overline{R^X}$ with respect to $u$ behaves as a Gaussian-like function. Thus, we can observe that:

\begin{enumerate}
\item[a)] the derivatives become more pronounced as $K$ increases. It is possible to establish a 
power law to describe this behavior. Then, 

\begin{equation}
\left( \frac{\partial R^X}{\partial u} \right)_{max} = B K^{\rho}.
\label{sca1}
\end{equation}

In a log–log scale the points are very well correlated by a linear function, as in
equation (\ref{sca1}), the numerical values of the fitting parameters being very similar for the
three criteria used here.

\item[b)] the derivatives are narrowed upon increasing $K$. 
This behaviour can also be described by a power law according to:

\begin{equation}
\Delta^X = C K^{-\lambda}.
\label{sca2}
\end{equation}

\noindent where $\Delta^X$ is the standard deviation of $\left( \frac{\partial R^X}{\partial u} \right)$ 
for each curve. Thus, the standard deviation
of each derivative versus $K$ when plotted in a log–log scale is very well correlated by a
linear function (not shown here), with the fitting parameter $\lambda =4.18 \pm 0.02$ for $A$, $U$ and
$I$ criteria.
\end{enumerate}

According to the above equations, a metric factor might be included in the scaling function,
equation (\ref{functionR}), in order to collapse all the curves in figure $3 (b)$ onto a single one. 
Following \cite{Privman1}, in figure $3 (c)$ we plot the probability $R^X_L$
as a function of the argument $u'= \left( \theta - \theta_c \right) L^{1/\nu} K^{\lambda}$.
As is clearly observed, all the curves collapse onto a single one. It is remarkable that more
than $2 \times 10^3$ points are included in the collapsing curve. The metric factor introduced here,
$K^{\lambda}$, gives an additional indication for the numerical value of the exponent $\lambda$ obtained in
equation (\ref{sca2}).

\section{Phase diagram and thermal finite-size scaling} \label{PD}

By using the scheme discussed above, the critical curve, $\theta_c$ versus $K$, separating the percolating
and non-percolating regions, is presented in figure $4$. In the studied range, three regimes can be
distinguished: $(i)$ for $K < -1.76$, $\theta_c$ remains constant as $K$ is decreased; 
$(ii)$ from $K \approx -1.76$ up to$K \approx 1.76$  
($K \approx -1.76$ ($K \approx 1.76$) being the reduced critical temperature for the
condensation (order–disorder) phase transition occurring in the system), $\theta_c$ increases linearly
with K and 
$(iii)$ for $K > 1.76$, $\theta_c$ remains constant as $K$ is increased. This behaviour can
be explained by simple geometrical arguments. Namely, lateral attractive interactions favour
the nucleation, which in turn increases the local connectivity and diminishes the percolating
fraction of occupied sites. In contrast, repulsive couplings avoid the occupation of nearest
neighbour sites, and consequently, increase the percolation threshold. In the limit cases, once
$K_c$ is reached, the adlayer does not vary significantly as $|K|$ is increased, and $\theta_c$ reaches its
saturation value. Thus, $\theta_c = 0.518$ for $K < -1.76$ and $\theta_c = 0.662$ for $K >1.76$.

As can be seen from the phase diagram, there exists an alternative route for determining
the critical curve. In fact, the surface coverage can be kept constant while the reduced
temperature, $K$, is varied. This procedure does not constitute the ‘standard’ technique for
calculating percolation features and it will be revised in detail in what follows. It is important
to emphasize that the idea of studying phase transitions upon varying a controlling parameter
and at the same time to keep constant the surface coverage is not new. In fact, this strategy has
been used in several works related with either kinetic (for example, in combination with the
well-known ZGB model, etc) \cite{Ziff2, Albano1, Albano2} or equilibrium phase transitions 
\cite{patryk}. However, to the
best of our knowledge this study has not been done for the present model despite its simplicity.

In figure $5$ typical curves of the probabilities $R^{I} _L(K)$ (squares), $R^{U} _L(K)$ (circles) and
$R^{A} _L(K)$ (triangles) are shown for two different values of coverage ($\theta = 0.5927$, full symbols
and $\theta = 0.638$, open symbols). For each case (fixed values of $\theta$ and a given criterion), just
three different lattice sizes are shown in the figure. The curves cross each other in a unique
point, $R^{X^*}$ , which depends on the chosen criterion. These points do not change their numerical
values regardless of the used coverage. 
These findings encourage for considering a finite-size
scaling analysis with the temperature as independent variable by following the same rules as
in section \ref{fss} where the percolation probability was calculated as a function of coverage. 
Thus, for each curve, $K_c ^X(L)$ is determined by least mean-square fitting. An extrapolation of 
$K_c ^X(L)$ towards the limit $L \to \infty$ by using the scaling hypothesis can be argued as

\begin{equation}
K_c ^X(L)=K_c(\infty)+ C^X L^{-1/\nu}
\label{extrapolation01}
\end{equation}

where $\nu$ is the critical exponent associated with the correlation length and $C^X$ is a constant for
each criterion used. 
Equation (\ref{extrapolation01}) is supported by numerical results as is shown in figure $6$ for
surface coverage $\theta = 0.626$ as an illustrative example. 
Thus, all the curves (different criteria) are well correlated by a linear function with a 
quite similar value for the ordinate in the limit $L\to \infty$. 
It is important to emphasize that $K_c ^A(L)$ is almost independent of the lattice size.
Therefore, the crossing points (figure $5$) are located at very well-defined values in the $K$-axes
determining the critical percolation threshold for each coverage.

By using standard finite-size scaling, it is possible to determine the critical exponent $\nu$
which results to be equal to $\nu = 4/3$ regardless of the value of $\theta$ considered as both 
(a) in the case of random percolation and (b) in section \ref{fss}.

As a consequence of the above results, it is possible to collapse all the curves in figure $5$
onto a single one for each coverage when $R^X _L (K)$ is plotted as a function of 
$z= \left( K - K_c \right) L^{1/\nu}$.
As an example, we plot $R^X _L$ as a function of $z$ for the surface coverage $\theta=0.638$ in
figure $7$, giving an additional indication for the calculated numerical value of the critical
exponent $\nu$. The same procedure described above was realized for different concentrations.
This fact demonstrates that the scaling function depends not only on the variable $K$ but also
on the coverage.

Thus, for determining the dependence of $\overline{R^X}$ with $K$, the main features of the collapsing
data must be considered: the curves become steeper upon increasing the coverage. In fact,
the derivative of the function $\overline{R^X}$ with respect to $z$ behaves as a Gaussian-like function. Thus,
we can observe the following facts: the derivatives (a) become more pronounced and (b) are
narrowed as the coverage increases. The latter is the most important finding where one wishes
to find the appropriate parametric factors in order to collapse all the curves, for different values
of $\theta$, into a single universal curve for each criterion. Such behaviour can also be described by
a power law according to

\begin{equation}
\Delta^X_L \propto  \theta^{-\Lambda}.
\label{sca4}
\end{equation}

\noindent where $\Delta^X_L$ is the standard deviation of 
$\left( \frac{\partial \overline{R^X}}{\partial z} \right)$. 

According to the above equations, a metric factor might be included in the scaling function
in order to collapse all the curves onto a single one for each criterion. 
In figure $8$ we plot the probability $R^X_L (K)$ as a function of the argument 
$z'= \left( K - K_c \right) L^{1/\nu} \theta^{\Lambda}$.
As is clearly observed, all the curves collapse onto a single one for each used criterion. 
This fact allows us to determine the numerical value of the parameter $\Lambda$ 
which results to be equal to $\Lambda =1.45 \pm 0.05$ regardless of the criterion used. 
The curves nicely collapse in the close vicinity of $z' = 0$
(close to the critical point) and a tiny deviation is observed as $|z'|$ increases. The scaling
analysis given above should be rigorously valid only for sufficiently large $L$ and from $T$ in the
asymptotic critical regime. However, it can be applied to the entire range of $L$ and $T$ if the data
fall in the ‘domain of attraction’ of a simple fixed point characterizing only one universality
class of critical phenomena.

It is remarkable that more than $2 \times 10^3$ points are included in the collapsing curve. The
metric factor introduced here, $\theta^{\Lambda}$, gives an additional indication for the numerical value of the
exponent $\Lambda$ obtained by using equation (\ref{sca4}).

\section{Conclusions}\label{C}

We presented a model to investigate the process of adsorption of interacting monomers on
a square lattice and studied the percolating properties of the adsorbed phase. By using
Monte Carlo simulation and finite-size scaling theory, we obtained the percolation thresholds
for different values of concentration and temperature. From this analysis, a critical curve
in the $\theta – T$ space was addressed. The coexistence line, separating the percolating and nonpercolating
regions, is characterized by three regimes: 
(1) for $K <-1.76$, $\theta_c$ remains constant ($\theta_c = 0.518$) as $K$ is decreased; 
(2) from $K \approx -1.76$ up to $K \approx 1.76$, $\theta_c$ increases almost linearly with $K$ 
and (3) for $K > 1.76$, $\theta_c$ remains constant ($\theta_c = 0.661$) as $K$ is increased.

Each point in the critical curve was corroborated by following an alternative scheme: the
surface coverage can be kept constant while the reduced temperature, $K$, is varied. This study
does not constitute the ‘standard’ procedure for calculating percolation properties. The results
in this paper show that the new technique seems to be a promising method for describing the
percolation behaviour of adlayer at equilibrium or what we called thermal percolation.

In all considered cases, the finite-size scaling study indicates that the model belongs to
the universality class of the random percolation model.

\section{Aknowledgments}
\label{sec:aknow}

This work was supported in part by CONICET (Argentina), FUNDACI\'ON ANTORCHAS
(Argentina) and the Universidad Nacional de San Luis (Argentina) under project 322000. One
of the authors (AJRP) is grateful to the Departamento de Qu\'imica, Universidad Aut\'onoma
Metropolitana-Iztapalapa (M\'exico, DF) for its hospitality during the time this manuscript was
prepared.

\parindent=0pt
\section*{Figure Captions}

\cap{1}{Fraction of percolating lattices as a function of the surface coverage. Different criteria are used for establishing the spanning cluster, namely, $R^{U} _L(\theta)$ the probability of finding either a rightward {\bf or} a downward percolating cluster (circles); $R^{I} _L(p)$ the probability that we find a cluster which percolates both in a rightward {\bf and} in a downward direction (squares); $R^{A} _L(p) \equiv \frac{1}{2} \left[R^{R} _L(p)+R^{D} _L(p) \right] \equiv \frac{1}{2} \left[R^{I} _L(p)+R^{U} _L(p) \right]$ (triangles). Three different values of $K$ are used as it is indicated. Horizontal dashed lines show the $R^{X^*}$ universal points. Vertical dashed lines denote the percolation threshold, $\theta_c$ in the thermodynamic limit $L \to \infty$. The error bars are smaller than the symbol size.}

\cap{2}{Extrapolation of $\theta_c (k)$ towards the thermodynamic limit
according to the theoretical prediction given by eq.(\ref{extrapolation}). Squares, triangles and circles denote the values of $\theta_c(k)$ obtained by using the criteria $I$, $A$ and $U$, respectively. Different values of $K$ are presented as indicated. The error bars are smaller than the symbol size.}

\cap{3}{$(a)$ Collapsing plot of the curves for the fraction of percolating samples as a function of $u$ For 
the case $K=0$. Each symbol denotes a different value of $L$ ($L=16, 32, 48, 64, 80, 96$ and $112$).
Each of the solid lines which are simply a guide for the eye represents one of the criteria $U$, $A$ and $I$, 
discussed in the text. $(b)$ The same as in (a) for different values of $K$ as indicated. The lines are just 
representative curves for the different criteria. $(c)$ The probability $R_L^X$ as a function of the argument $u'= \left( \theta - \theta_c \right) L^{1/\nu} K^{\lambda}$, where the metric factor $K^{\lambda}$ is included in order to collapse all the curves in Fig. $3a$ onto a single one for each criterion.}

\cap{4}{Phase diagram, $\theta_c$ vs. $K$, which shows the curve separating the percolating and not percolating 
regions. Vertical dashed lines at $K= -1.76$ and $K= 1.76$ denote the reduced critical temperature for the 
phase transition occurring in the adlayer phase for attractive and repulsive interacting particles 
respectively. Horizontal dashed lines at $\theta_c = 0.518$ and $\theta_c = 0.662$ are the critical coverage 
at saturation regime for $K < -1.76$ and $K > 1.76$ respectively. The error bars are smaller than 
the symbol size.}

\cap{5}{Typical curves of the probabilities $R_L^I (K)$ (squares), $R_L^U (K)$ (circles) and $R_L^I (K)$ 
(triangles) are shown for two different values of coverage. $\theta = 0.5927$ corresponds to full symbols 
while $\theta = 0.638$ is denoted by open symbols. For each case (fixed values of $\theta$ and a given 
criterion), just three different lattice sites are shown in the figure ($L = 32$, $48$ and $64$). The 
universla points $R^{X^{\ast}}$ are denoted by horizontal dashed lines while the critical temperatures in each 
case are represented by vertical dashed lines. The error bars are smaller than the symbol size.}

\cap{6}{Extrapolation of $K_C^X (L)$ towards the thermodynamic limit according to the theoretical 
prediction given by equation \ref{extrapolation01} for surface coverage $\theta = 0.626$. Squares, 
triangles and circles denote the values of $K_c(L)$ obtained by using the criteria $I$, $A$ and $U$ 
respectively. The numerical value of $K_C(\infty)$ is indicated in excellent agreement with the phase 
diagram shown in figure $4$. The error bars are smaller than the symbol size.}

\cap{7}{Collapsing plot of the curves for the fraction of percolating samples $R_L^X (K)$ as a function 
of $z = (K - K_C) L^{1/\nu}$ for the case $\theta = 0.638$. Each symbol denotes a different value of $L$ 
($L= 16, 32, 48, 64, 80, 96$ and $112$). Each solid line, which is simply a guide for the eye, represents 
one of the criteria, $U$, $A$ and $I$, discussed in the text.}

\cap{8}{The probability $R_L^X$ as a funtion of the argument $z' = (K - K_C) L^{1/\nu} \theta^{\Lambda}$, 
where the metric factor $\theta^{\Lambda}$ is included in order to collapse all the curves onto a single 
one for each criterion.}

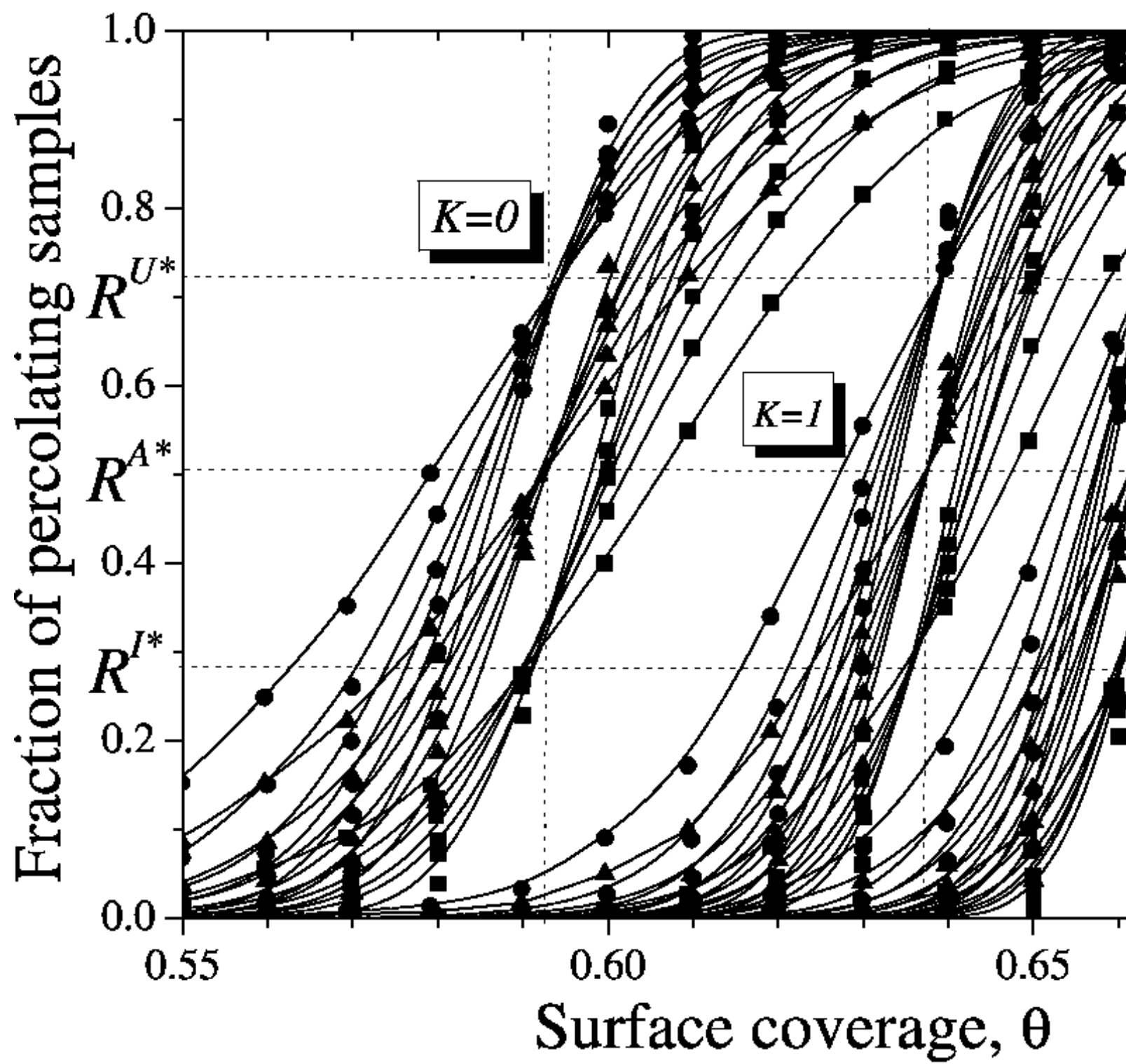
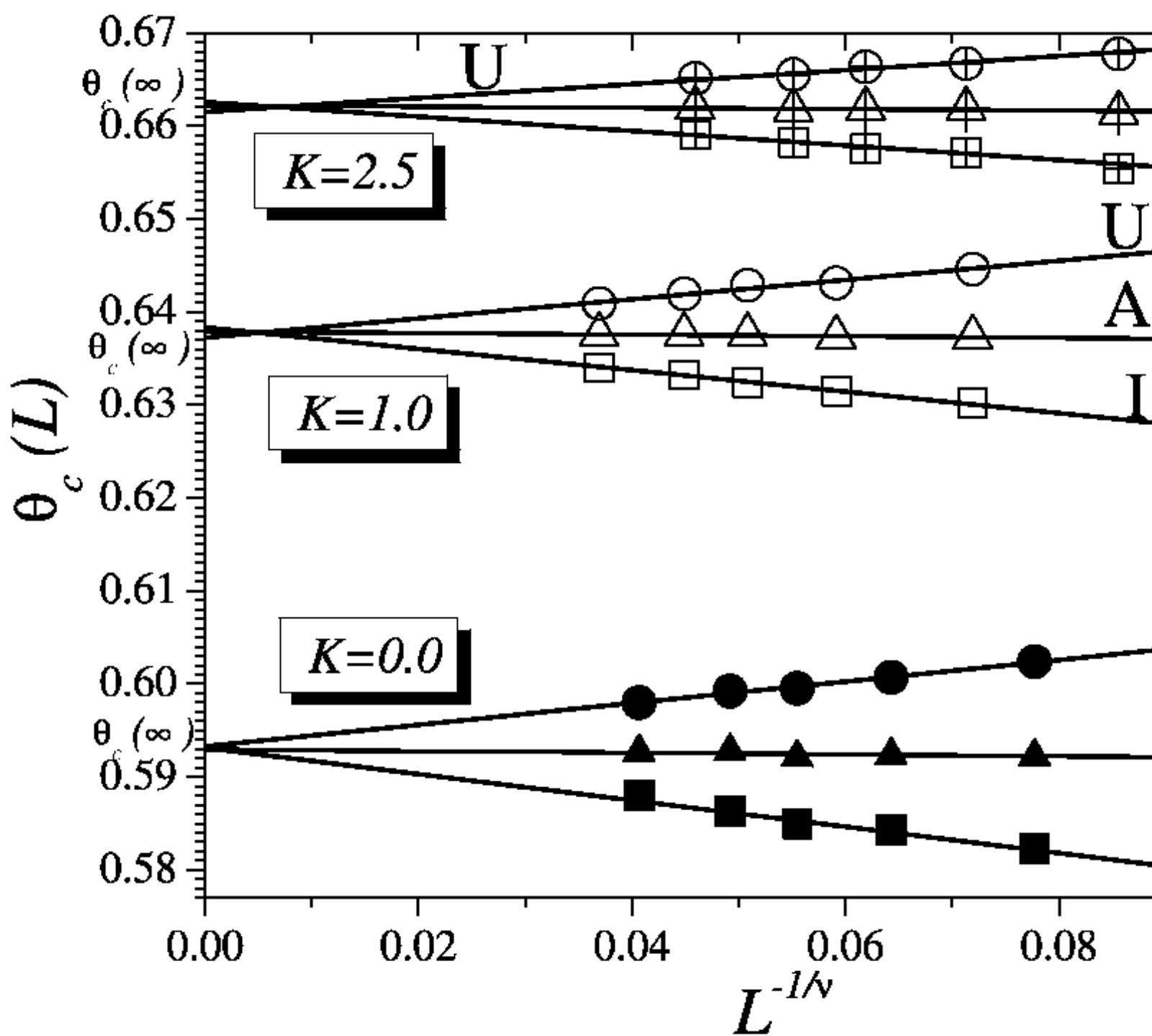
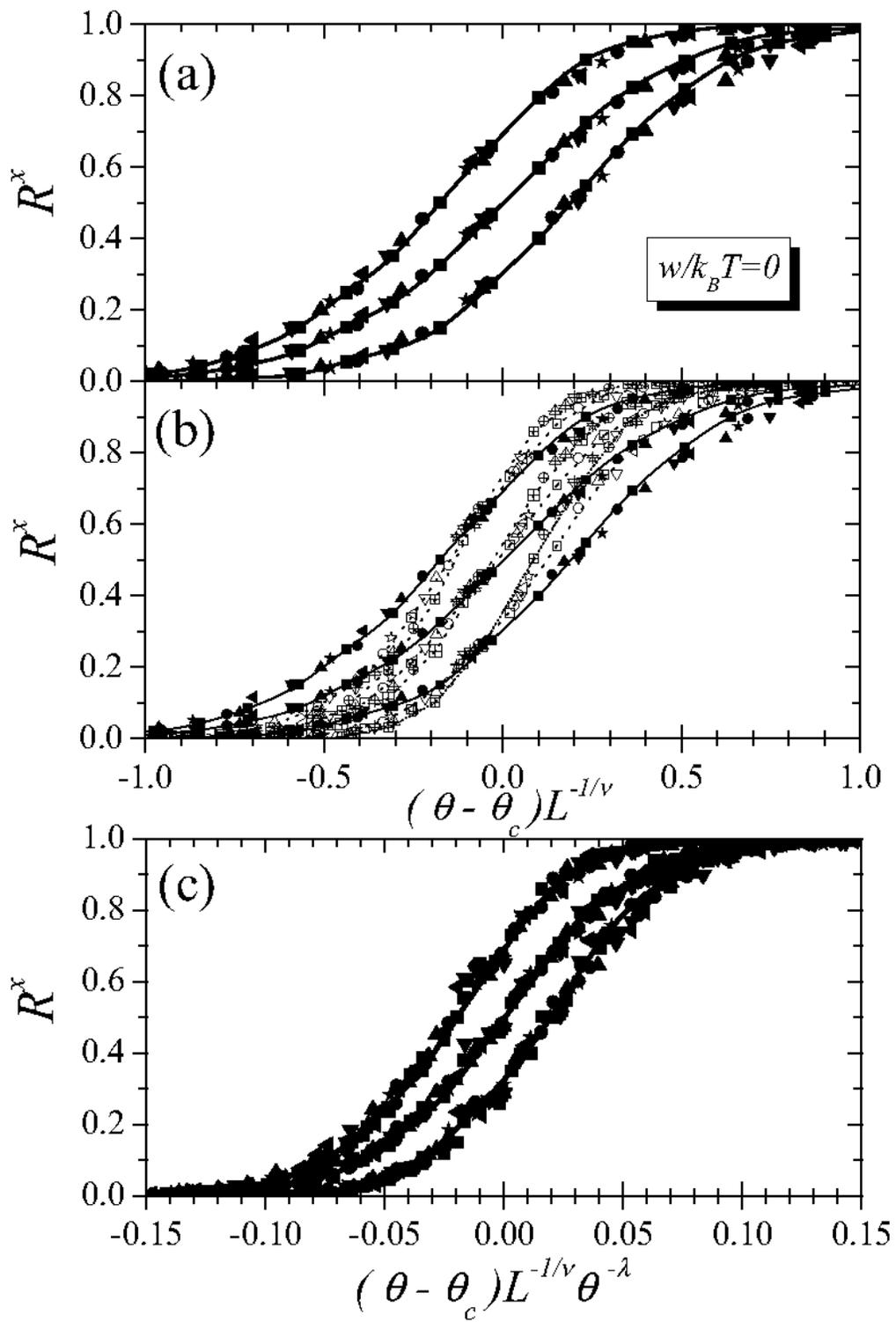
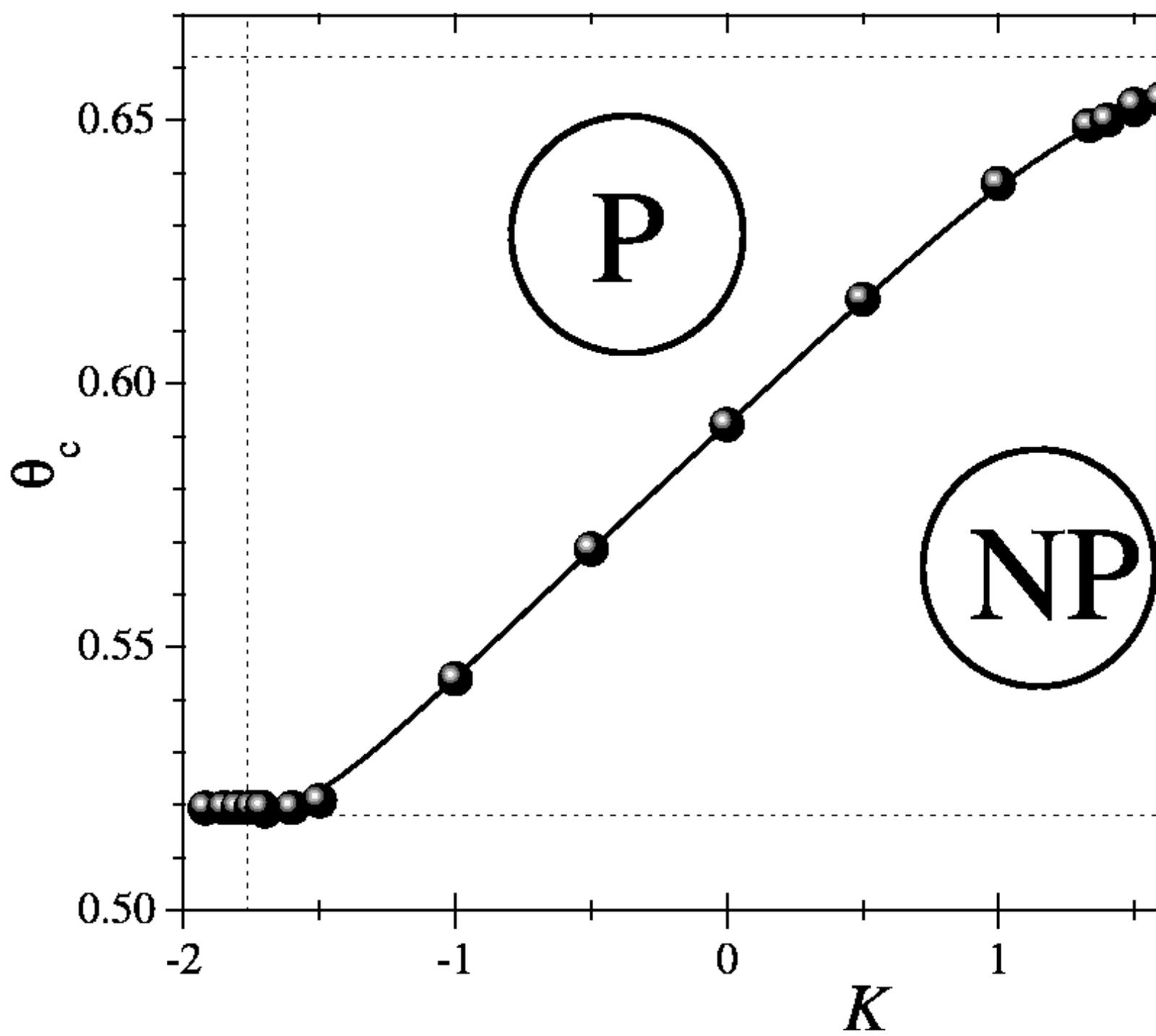
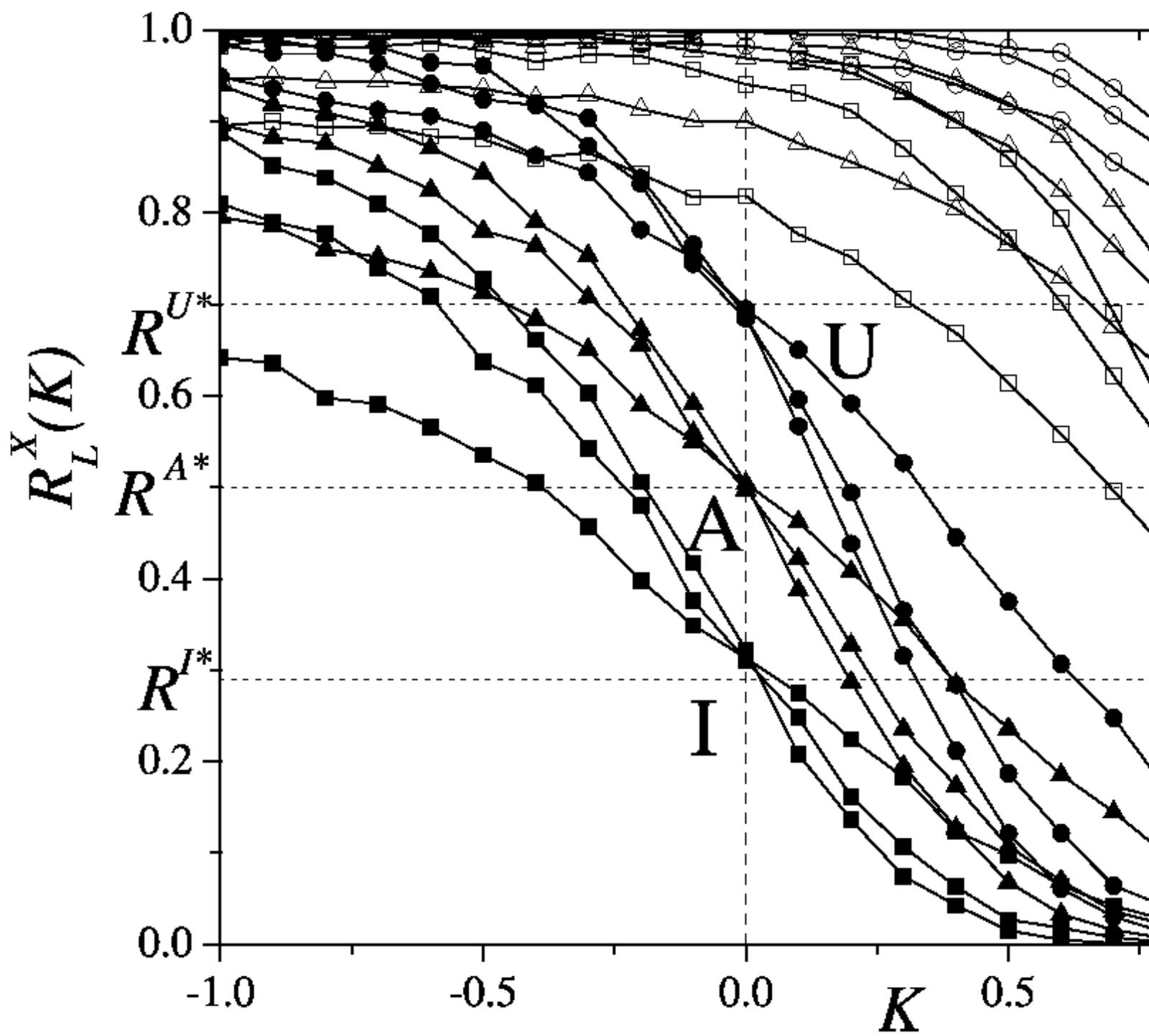
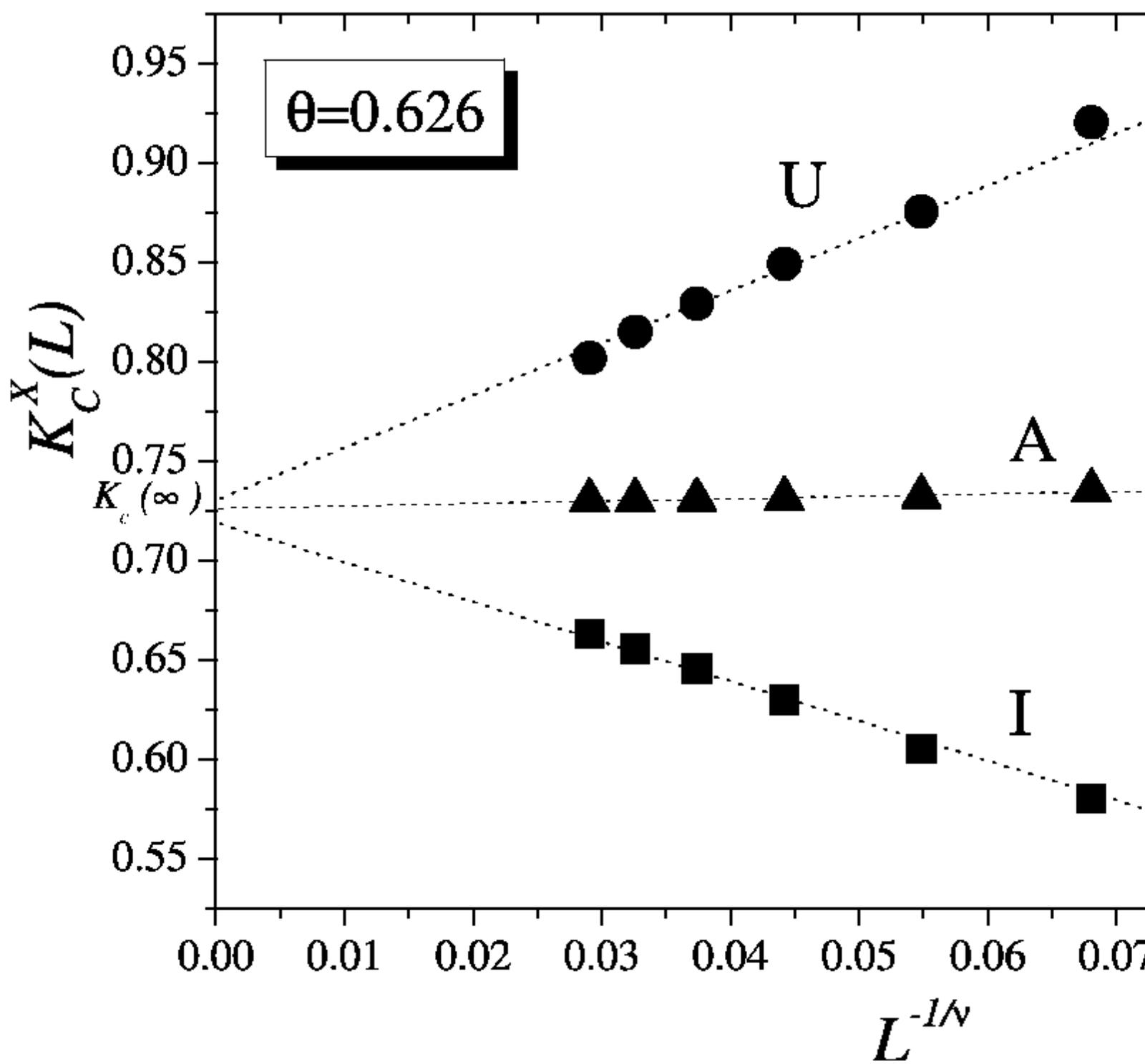
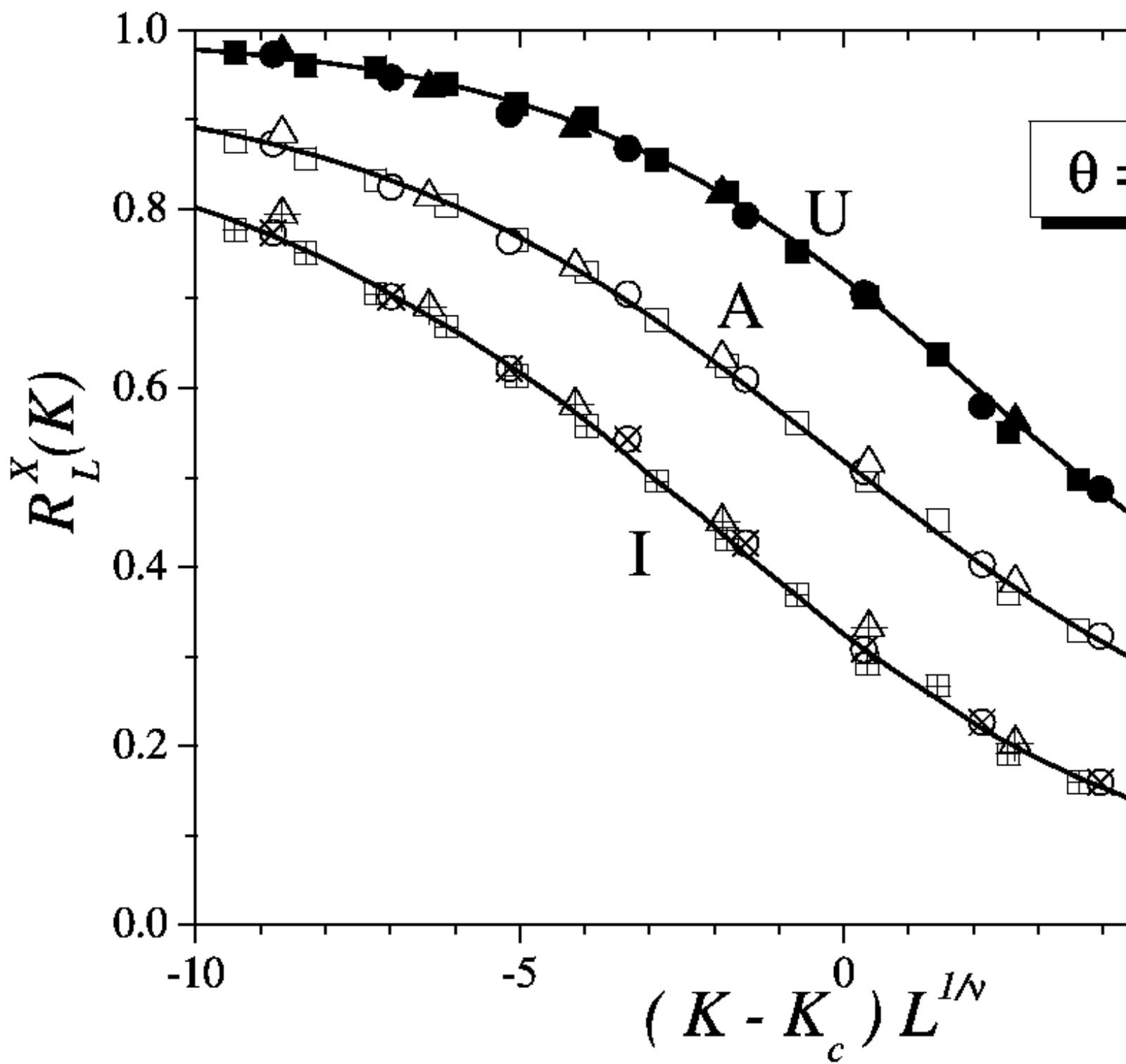
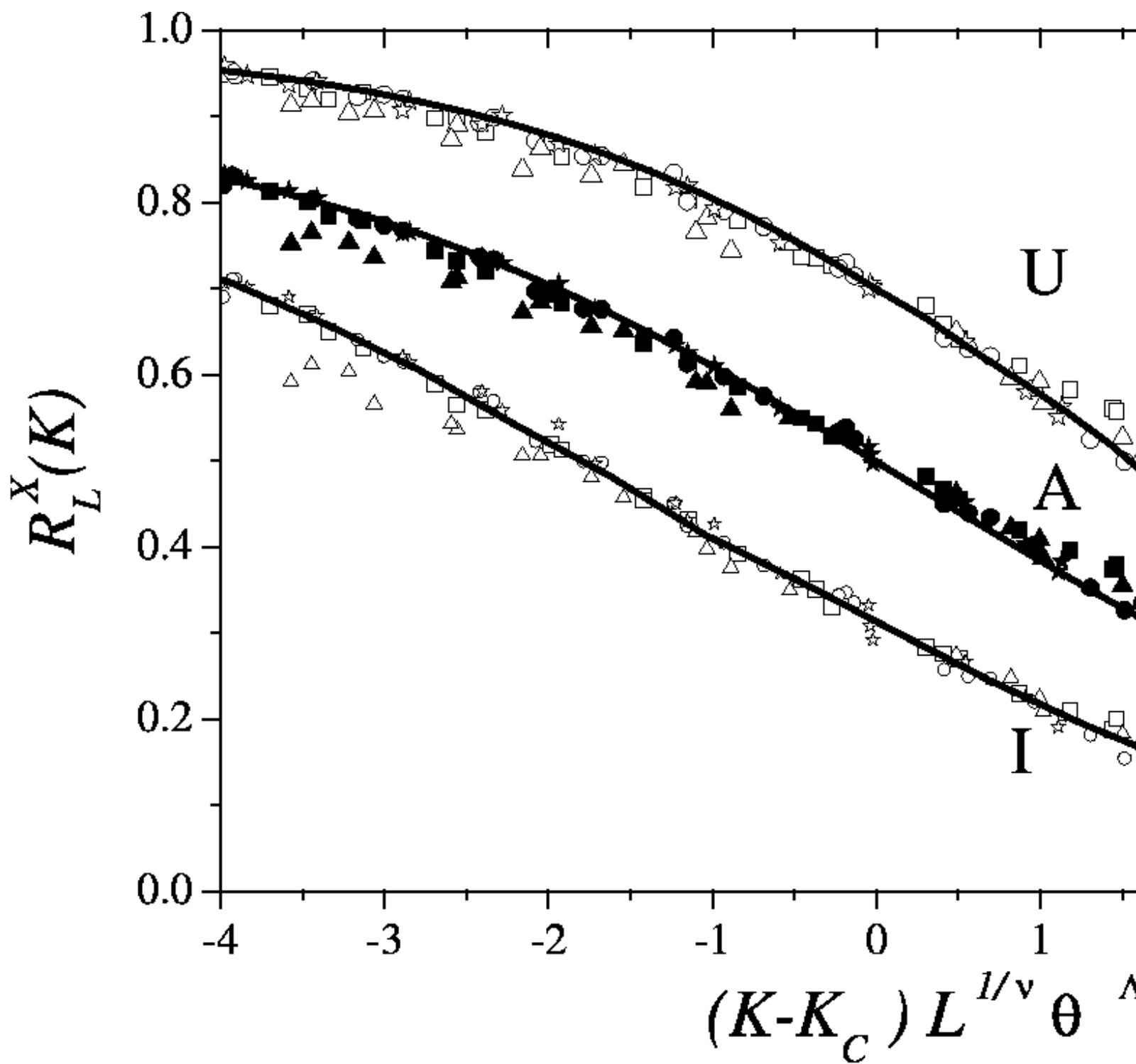
\end{document}